\documentclass[aps,prd,superscriptaddress,twocolumn]{revtex4-1}
\usepackage{graphicx,times,xcolor,subfigure}
\usepackage{amsmath,amssymb,amsthm}
\usepackage{hyperref,cleveref}
\pdfoutput=1
\usepackage{cleveref}

\newcommand{\tr}[1]{\hbox{Tr}\left[#1\right]}
\newcommand{\ptr}[2]{\hbox{Tr}_{#1}\left[#2\right]}
\begin{document}

\title{Discrimination of Ohmic thermal baths by quantum dephasing probes}
\author{Alessandro Candeloro}
\email{alessandro.candeloro@unimi.it}
\affiliation{Quantum Technology Lab, Dipartimento di Fisica 
{\em Aldo Pontremoli}, Universit\`a degli Studi di Milano, I - 20133, 
Milano, Italy.}
\author{Matteo G.A. Paris}
\email{matteo.paris@fisica.unimi.it}
\affiliation{Quantum Technology Lab, Dipartimento di Fisica 
{\em Aldo Pontremoli}, Universit\`a degli Studi di Milano, I - 20133, 
Milano, Italy.}
\begin{abstract}
We address the discrimination of structured baths at different temperatures by 
dephasing quantum probes. We derive the exact reduced dynamics and evaluate the 
minimum error probability  achievable by three different kinds of quantum probes, 
namely a qubit, a qutrit and a quantum register made of two qubits. Our results 
indicate that dephasing quantum probes are useful in discriminating low values 
of temperature, and that lower probabilities of error are achieved for 
{intermediate values} of the interaction time. A qutrit probe outperforms a 
qubit one in the discrimination task, whereas a register made of two qubits 
does not offer any advantage compared to two single qubits used sequentially. 
\end{abstract}
\date{\today}
\maketitle
\section{Introduction}
Thermometry  is about  measuring  the thermodynamic 
temperature of a system. In classical thermodynamics,  thermometry is based on 
the zeroth principle, i.e. it relies on the achievable equilibrium 
between the system and a probe with a much smaller heat capacity. In quantum mechanics, 
temperature is not an observable in a strict sense. Rather, it is a parameter on which the 
state of a quantum system may depend on. For this very reason, direct measurement of temperature 
is not available, and one should resort to indirect measurement procedures. During the last 
decade, quantum thermometric strategies have emerged \cite{mehboudi2019thermometry,de2018quantum,potts2019fundamental,landb}, 
which are mostly based on using an 
external quantum probes interacting with the system under investigation, with the 
assumption that the interaction between the probe and the system does not change the 
temperature of the latter. Those strategies, usually termed {\em quantum probing} schemes, 
are not based on the zeroth principle, but rather on engineering of the interaction 
Hamiltonian, which is exploited to imprint the temperature of the system on the 
quantum state of the probe. As a matter of fact, quantum probing 
exploits the inherent fragility of quantum systems against decoherence, turning 
it into a resource to realize highly sensitive metrological schemes. 
\par
In the recent years, temperature estimation by quantum probes received much attention
\cite{bru06,sta10,bru11,bru12,mar13,hig13,meh15,jar15,adp15,bb20}, often using the tools 
offered by quantum estimation theory. The optimal sensitivity in temperature estimation 
has been studied for $N$-dimensional quantum probes \cite{correa2015individual} and, more 
recently, the efficiency of infinite-dimensional quantum probes have been also 
investigated \cite{mancino2020non}. The ultimate quantum limits to thermometric precision
has been addressed \cite{landb}, as well as the use of out-of-equilibrium 
quantum thermal machine has been suggested for temperature estimation 
\cite{hofer2017quantum}. Quantum thermometry 
by dephasing has been also addressed in details and, in particular, the performance of 
single qubit probes \cite{razavian2019quantum} and of quantum registers made of two 
qubits \cite{gebbia2020two} have been explored.
\par
\begin{figure}[h!]
\includegraphics[width=0.9\columnwidth]{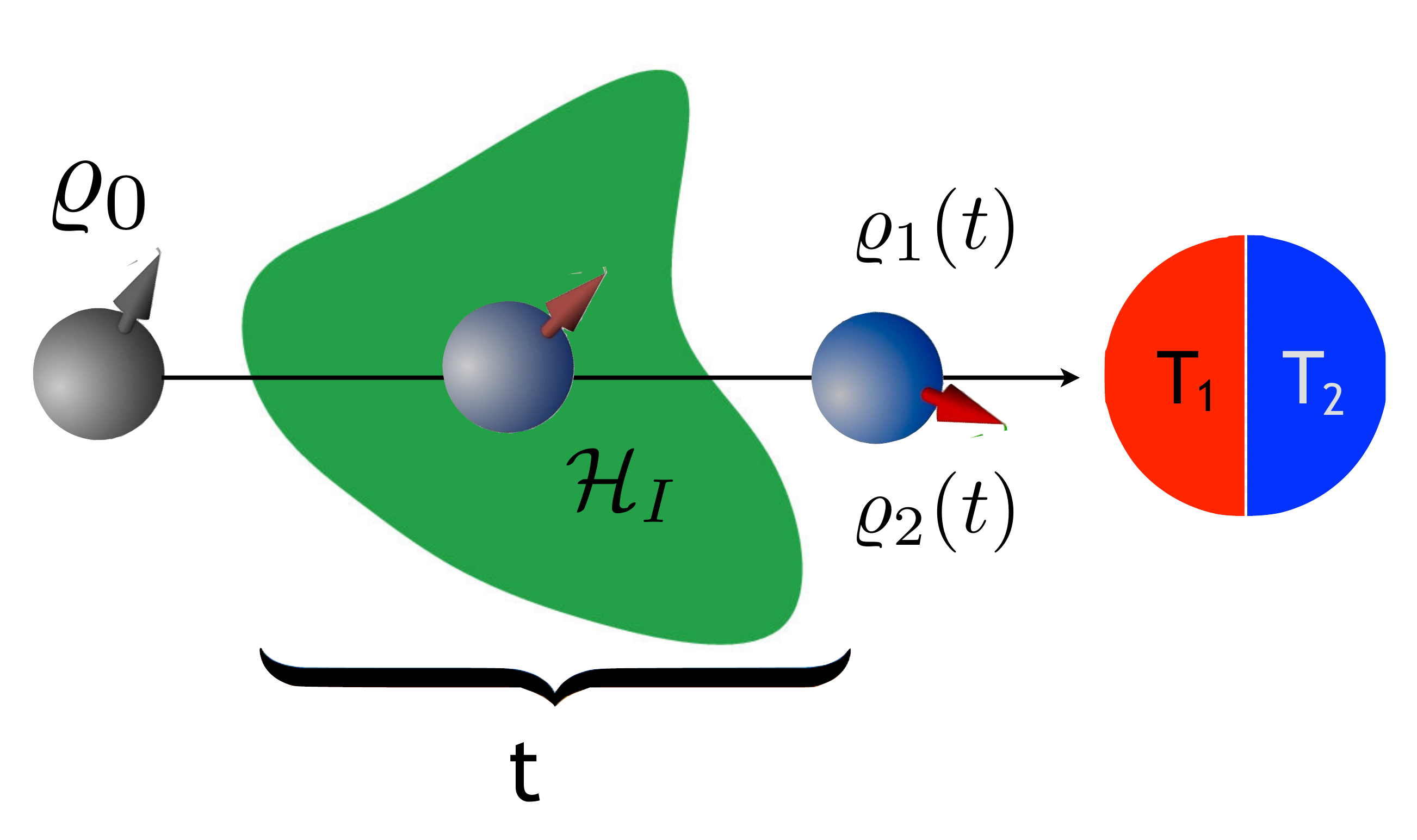}
\caption{Discrimination of temperatures by quantum probes. A quantum 
system prepared in a known state $\varrho_0$ is let to interact with 
a thermal bath for a time $t$ and then measured in order to discriminate 
whether the state is $\varrho_1 (t)$ or $\varrho_2 (t)$, i.e. to
infer whether the temperature of the bath is $T_1$ or $T_2$. After 
choosing a suitable interaction Hamiltonian ${\cal H}_I$, the scheme 
may be optimized over the initial preparation of the probe, 
and the value of the interaction time $t$. \label{f:f1_sch}}
\end{figure}

As a matter of fact, less attention has been devoted to estimation of a discrete 
sets of temperature values, i.e. to temperature discrimination. The problem  is that 
of telling apart thermal baths with different temperatures, assuming that the possible 
values of temperature belong to a discrete sets $\{T_1,T_2,...\}$ and are known in 
advance (see Fig. \ref{f:f1_sch} for a pictorial description of the measurement scheme). 
\par
\par
In this framework, a single qubit has been suggested \cite{jevtic2015single} 
as an out-of-equilibrium probe to discriminate two thermal baths and, more recently, 
the discrimination between baths with different temperatures or statistical 
properties has been addressed \cite{gianani2020discrimination}, assuming that 
the quantum probe undergoes Markovian dynamics. In this paper, we extend these 
studies to more general quantum probes and taking into account the spectral 
structure of the bath.  In particular, we assume a dephasing interaction between 
the probe and the bath, and derive the exact reduced evolution of the quantum 
probe. Then, we study the discrimination performance of our scheme for different 
kinds of Ohmic-like environments and for  different quantum probes. In order to
provide a benchmark, we first analyze discrimination by quantum probes at 
equilibrium, and then address the out-of-equilibrium case, looking for the optimal 
interaction time, leading to the smallest error probability. Our results clearly indicate 
that dephasing quantum probes are useful in discriminating low values of temperature, and
that lower probabilities of error are achieved for {intermediate values} of 
the evolution time, i.e. for out-of-equilibrium quantum probes \cite{shpa}.
\par
The paper is structured as follows. In the next Section, we review some elements 
of quantum discrimination theory and establish notation. In Section \ref{sec:thermaleq}, 
we analyze discrimination of thermal baths by quantum probes at equilibrium. Besides 
being of interest in their own, the results of this Section serve as a benchmark 
to assess the performance of out-of-equilibrium quantum probes, which are analyzed 
in details in Section \ref{sec:thermalnoneq}. Section \ref{subsec:comparison} 
closes the paper with some concluding remarks, whereas few more details about 
the reduced dynamics of the quantum probes are reported in the Appendix.
\section{\label{sec:discri} The quantum discrimination problem}
In several problems of interest in quantum technology, an observer 
should discriminate between two or more quantum states. However,
quantum states are not observable and this operation cannot be 
carried out directly. Furthermore, distinct states may have finite 
overlap, and  there is no way to distinguish them with certainty \cite{chefles2000quantum}. The main consequence is 
that a correct discrimination among a generic set of quantum states 
is not always possible, and an intrinsic error in the process occurs. 
Many strategies for optimal discrimination of quantum state\cite{Bergou2004,barnett2009quantum,bae2015quantum} have been suggested, each of
them tailored to specific purpose. In this paper, we are going to use the minimum 
error discrimination strategy, which we briefly review in the following. 
\par
Let us consider the problem of binary discrimination between two quantum 
states $\rho_1$ and $\rho_2$ which are known in advance and occur  
with an {\em priori} probability $\{z_k\}$, $k=1,2$. Given a probability operator-valued measure (POVM) $\{\Pi_1,\Pi_2\}$, the quantity $\tr{\Pi_j \rho_j}$ represents the probability of correctly infer the state $\rho_j$ by implementing the POVM.  In order to optimize the discrimination, the POVM 
must be chosen to minimize the overall 
probability of error, i.e. 
\begin{equation}
    p_{e} = 1- \sum_{j=1}^{2} z_j \tr{\Pi_j \rho_j},
\end{equation}
Since $z_1 + z_2 = 1$ and $\Pi_1 + \Pi_2 = \mathbb{I}$, $p_{e}$ 
may be rewritten as $p_{e} = p_1 + \tr{ \Lambda \Pi_1} = 
p_2 - \tr{ \Lambda \Pi_2}$
where the Hermitian operator $\Lambda$ is defined as
\begin{equation}
    \Lambda = z_2 \rho_2 - z_1 \rho_2.
    \label{eq:Lambda}
\end{equation}
Using the spectral decomposition $\Lambda = \sum_k \lambda_k \vert 
\psi_k\rangle\langle\psi_k \vert $, the minimum probability of error 
may be written in terms of the so-called Helstrom bound \cite{helstrom1969quantum,bergou2007quantum}
\begin{equation}
    p_{e} = \frac{1}{2} \left(1-\sum_k \vert \lambda_k \vert \right) = \frac{1}{2} \left(1-\tr{\vert \Lambda \vert}\right).
    \label{eq:perrautoval}
\end{equation}
Using the distance norm \cite{nielsen2002quantum} we can interpret the result from a geometrical point of view. Since $\tr{\vert \Lambda \vert} = \tr{\vert z_2 \rho_2 - z_1 \rho_1\vert} = \Vert z_2 \rho_2 -z_1 \rho_1 \Vert_1$, if the occurrence probabilities of $\rho_1$ and $\rho_2$ are the same, we obtain
\begin{equation}
     p_{e} =  \frac{1}{2} \left[1-D(\rho_1,\rho_2) \right]\,,
     \label{eq:perr}
 \end{equation} 
 where $D(\rho_1,\rho_2) = \frac12 \Vert \rho_2 -\rho_1 \Vert_1 $ is the trace distance.
This result confirms our intuition that the less two states are distant, the larger is the probability of error in discriminating them. We also remind 
that the optimal POVM, for which the probability of error is minimized, 
is given in terms of the eigenprojectors of the operator $\Lambda$, as $\Pi_0 = \sum_{\lambda_k \leq 0}\vert 
\psi_k\rangle\langle\psi_k \vert $
\section{\label{sec:thermaleq} Quantum probes at thermal equilibrium}
Let us now turn to the main problem of the paper, i.e. to discriminate 
whether a thermal bath is at temperatures $T_1$ or $T_2$ by performing measurements on a quantum probe interacting with it. In this Section, 
we assume that the probe is at the equilibrium with the bath. We do 
not study how the probe reaches the equilibrium with the bath, and simply 
assume that after enough time the probe has reached such equilibrium.
In the next Section, we devote attention to the out-of-equilibrium case
and will introduce an interaction model.
\par
Let us consider quantum system governed by a bounded Hamiltonian 
$\mathcal{H}$ with an energy spectrum 
$\{\vert e_n\rangle, E_n\}_{n=0}^{N-1}$, then the 
equilibrium state of the probe is given by the Gibbs state
\begin{equation}
    \rho_{eq}(\beta) = \frac{1}{Z(\beta)}\sum_{n= 0}^{N-1}e^{-\beta E_n} 
    \vert e_n \rangle \langle e_n \vert
    \label{eq:equilbriumstate}
\end{equation}
where $Z(\beta)$ is the partition function $Z(\beta) = \sum_n e^{-\beta E_n}$ and $\beta = 1/ T$ 
(we set the Boltzmann constant to $1$ throughout the paper) is the inverse temperature 
of the heath bath. 
\par
Consider now the situation where we do not know in advance the temperature of the bath, 
but we know it must be $T_1$ or $T_2$. As a result, the thermal state will be different 
and our goal is to discuss the minimum probability of error in discriminating the two 
states $\rho_{eq}(\beta_1)$ and $\rho_{eq}(\beta_2)$.  From the previous Section, 
we know that the best measurement is given by the operator $\Lambda$ in \eqref{eq:Lambda}. 
In our case, since both states are diagonal in the energy eigenbasis of $\mathcal{H}$, 
the optimal measurement is an energy measurement. The probability of error in the 
discrimination is given by \eqref{eq:perr}, that is
\begin{equation}
    p_{e}^{eq}(\beta_1,\beta_2) = \frac{1}{2}-\frac{1}{4}\sum_{n = 0}^{N-1} \left\vert \frac{e^{-\beta_1E_n}}{Z(\beta_1)} - \frac{e^{-\beta_2E_n}}{Z(\beta_2)} \right\vert.
    \label{eq:peqerr}
\end{equation}
When one of the temperature is vanishing, say $T_2 = 0$ ($\beta_2 = + \infty$), the 
corresponding thermal probe collapses into the ground state $\vert e_0 \rangle 
\langle e_0 \vert$ and the probability of error becomes
\begin{align}
    p_{e}^{eq}(\beta_1,+\infty) & = \frac{1}{2} - \frac{1}{4} \left(\left\vert  \frac{e^{-\beta_1 E_0}}{Z(\beta_1)} - 1\right\vert + \sum_{n = 1}^{N-1} \left\vert\frac{e^{-\beta_1 E_n}}{Z(\beta_1)}\right\vert \right) \nonumber \\
    & = \frac{1}{2}\frac{e^{-\beta_1 E_0}}{Z(\beta_1)}
    \end{align}
In the opposite limit, i.e. when on of the two baths has a very large temperature, say $T_2$ is very large ($\beta_2 \to 0$) compared to the 
largest energy eigenvalue $\max_n\{E_n\}$, the corresponding thermal state 
approaches the equiprobable diagonal state $\rho_{eq}(0) = \mathbb{I}/N$. 
\begin{figure}[h!]
\centering
\includegraphics[width=0.8\columnwidth]{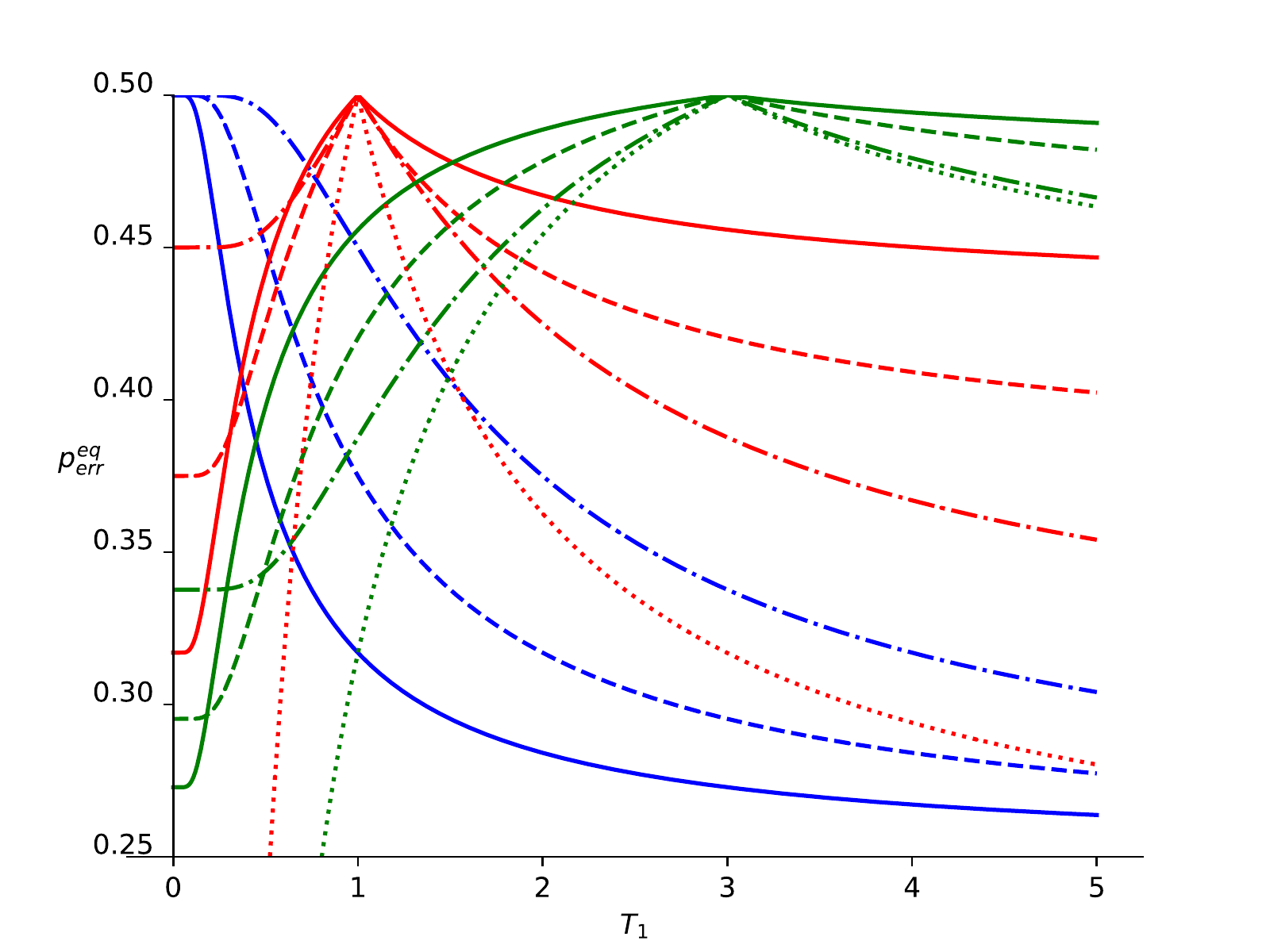}
\includegraphics[width=0.8\columnwidth]{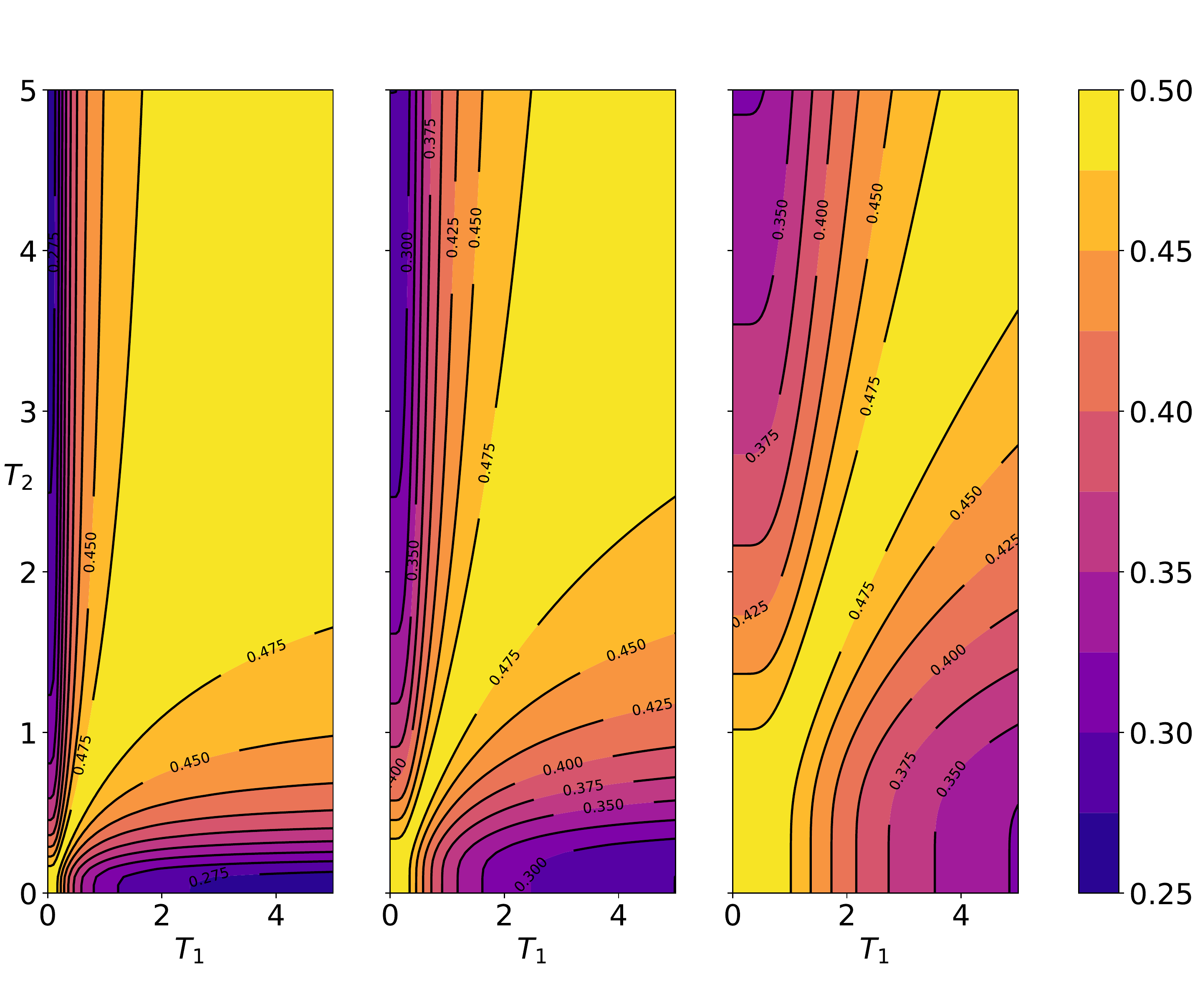}
\caption{Upper panel: the probability of error at equilibrium $p_{e}^{eq}(T_1,T_2)$ for a qubit probe \eqref{eq:perreqqubit} as a function 
of $T_1$ and for different values of $T_2$ and $\omega_0$. Blue line $T_2 =0$; Red line $T_2 = 1$; Green line $T_2 = 5$. Thick Line $\omega_0 = \log(3)/2$, Dashed Line $\omega_0 = \log(3)$, DotDashed Line $\omega_0 = 2 \log(3)$. Lighter Red (Green) is the approximated probability of error \eqref{eq:perreqlargeT} with $T_2 = 1$ ($T_2 = 5$) and $\omega_0 = \log(3)$. Lower panels: Contour plots of $p_{e}^{eq}$ as a function of $T_1$ and $T_2$. Left: $\omega_0= \log(3)/2$; middle: $\omega_0= \log(3)$; right: $\omega_0= 2\log(3)$.}
\label{f:f1eq}
\end{figure}
\par
In this limit, the partition function, up to terms $O(\beta^2)$, may be written as 
\begin{align}
    Z(\beta) &  \simeq \sum_{n=0}^{N-1} \left(1-\beta E_n\right) =  N\left(1-\beta \bar{E}\right)\,,
\end{align}
with $\bar{E} = \sum_{n=0}^{N-1} E_n/N$. The Boltzmann weight becomes
\begin{equation}
    \frac{e^{-\beta E_n}}{Z(\beta)} \simeq \frac{1}{N} - \frac{\beta}{N}\left(E_n-\bar{E}\right) \,,\end{equation}
and the probability of error is given by
\begin{align}
    p_{e}^{eq}(\beta_1,&\beta_2) = \frac{1}{2} \left(1 -   
 \frac{1}{2N}  \sum_{n=0}^{N-1} \left \vert \left(E_n-\bar{E}\right) \left(\beta_1- \beta_2\right) \right\vert  \right)
    \label{eq:perreqlargeT}
\end{align}
\par
For a two-dimensional (qubit) probe $d=2$, a closed formula for \eqref{eq:peqerr} may 
be easily  evaluated, obtaining
\begin{align}
    p_{e}^{eq}&(\beta_1,\beta_2) = \frac{1}{2}\left[ 1+ 
  \left|\tanh\left(\frac{\omega_0\beta_2}{2}\right)-\tanh\left(\frac{\omega_0\beta_1}{2}\right)\right|\right]\,.
    \label{eq:perreqqubit}
\end{align}
We will use this expression in the following Section, to compare performance of probes at 
equilibrium with that of out-of-equilibrium ones.
\par
In the upper panel of Fig. \ref{f:f1eq}, we show $p_{e}^{eq}(T_1,T_2)$ 
as a function of $T_1$ for a qubit system with frequency $\omega_0$, ($E_0 =-\omega_0/2$ and $E_1 = \omega_0/2$) and for different values of $T_2$. 
We see that the minimum of $p_{e}^{eq}(T_1,T_2)$ depends on the relative 
choice of $T_1$ and $T_2$. If $T_2 = 0$, the minimum is reached asymptotically for $T_1 \to +\infty$, and we know from previous considerations that the limiting values is equal to $1/4$. Instead, for $T_2 > 0$, we have two cases: if $T_2 \leq \omega_0\log(3)$, the minimum of $p_{e}^{eq}(T_1,T_2)$ is reached for $T_1 \to 0$, while if $T_2 \geq \omega_0/\log(3)$ then the minimum of $p_{e}^{eq}(T_1,T_2)$ is again obtained asymptotically for $T_1 \to +\infty$. In the lower panels of Fig. \ref{f:f1eq}, we show $p_{e}^{eq}(T_1,T_2)$ for the same qubit system as a function of $T_1$ and $T_2$ and for different values of $\omega_0$.  We may clearly 
see the symmetry between $T_1$ and $T_2$. We also notice that as $\omega_0$ grows, discrimination improves, especially in the high temperature regime. This can be understood from \eqref{eq:perreqlargeT}, since for larger $\omega_0$ the second term, which  is proportional to $\omega_0$ in the qubit case, is larger and thus $p_{e}^{eq}(T_1,T_2)$ is smaller.
\section{\label{sec:thermalnoneq}  Out-of-equilibrium dephasing quantum probes}
Let us now study how dephasing, out-of-equilibrium, quantum probes 
may be exploited in the  temperature discrimination problem. Here, the quantum probe is an open quantum system $S$ which effectively interacts with the reservoir, which is a thermal bath at temperature $T_1$ or $T_2$. We assume that the total Hamiltonian of the system is $\mathcal{H} = \mathcal{H}_0^S + \mathcal{H}_0^B + \mathcal{H}_I$, where the first term determines the free evolution of the system, the second the free evolution of the bath and the latter the interaction between the open quantum system and the reservoir. Before specifying the interaction model,let us discuss some general results about temperature discrimination, regardless 
of the system and of the interaction.
\par
To perform our discrimination task, we prepare our quantum probe in a certain state 
and then we let it interact with the bath. 
We assume that the bath is at equilibrium in a Gibbs state
\begin{equation}
   \nu_k = \frac{e^{-\beta_k\mathcal{H}_0^B}}{Z(\beta_k)}.
\end{equation}
where $\beta_k, k=1,2$ are two distinct inverse temperature.
Once fixed the probe state $\rho_S\equiv\rho_S(0)$ at time $t=0$ 
and the environment state $\nu_k$, the evolution of 
the initially factorized total system $\rho_S
\otimes \nu_k$ is determined by a 
completely-positive trace-preserving (CPT) map $\Phi^k_t$. The 
state of the system at time $t$ will be 
\begin{align}
    \rho_{Sk}(t) =  \Phi^k_{t}[\rho_S] = \ptr{E}{U(t)\, \rho_S \otimes \nu_k\, U^\dag(t)}.
\end{align}
The two baths at different temperature define two different CPT maps, and we are going 
to see that the distance between these two different maps, defined in the last equation, 
has an upper bound which do not depend on the nature of the probe.
The probability of incorrectly discriminate the two states originating from the interaction 
with the two baths is \eqref{eq:perr}, and it depends on the trace distance 
$D(\rho_{S1}(t),\rho_{S2}(t))$. Since the trace distance is contractive under 
the action of trace-preverving map, and invariant under unitary transformations, \cite{nielsen2002quantum,breuer2016colloquium}, we have
\begin{align}
    D(\rho_{S1}(t), \rho_{S2}(t)) & = D\Big(\Phi^1_t[\rho_S],\Phi^{2}_t[\rho_S]\Big) 
    \notag \\ 
    & \leq D(\rho_S\otimes \nu_1, \rho_S\otimes \nu_2)
        \notag \\
        & = D(\nu_1,\nu_2)\,,
\end{align}
where the last equality is due to the fact that the state of quantum probe at time $t=0$ 
is fixed, regardless of the temperature, and thanks to the additivity under tensor 
products of the trace distance. 
This is an upper bound on the maximum distance between two states evolving under the same reduced dynamics with two baths at $T_1$ and $T_2$. Moreover, this bound depends only on the nature of the bath (namely its Hamiltonian $\mathcal{H}_0^B$) and on the temperatures to be discriminated. The upper bound translates into a lower bound on the probability of error \eqref{eq:perr}, that is
\begin{equation}
    p_{e}^{neq}(T_1,T_2) \geq \frac{1}{2}\left[1- D(\nu_1,\nu_2)\right]\,.
\end{equation}
This bounds may be useful in dealing with finite size environment, whereas in the 
thermodynamical limit the {\em orthogonality catastrophe} is likely to make it useless 
\cite{oc0,oc1}. 
\subsection{Dephasing model}
In this section we introduce a (pure) dephasing model that regulate the probe-environment 
interaction by generalizing the qubit model studied in \cite{breuer2002theory}. The 
full dynamics is generated by the Hamiltonian
\begin{equation}
    \mathcal{H}_T = \mathcal{H}_0 + \mathcal{H}_I,
    \label{eq:fulldynamics}
\end{equation}
where $\mathcal{H}_0 = \mathcal{H}_0^S + \mathcal{H}_0^B$ determines the free evolution 
of the probe and the bath, whereas $\mathcal{H}_I$ describes the interaction. Since we 
are going to consider quantum probes with a discrete energy spectrum, we may introduce 
an energy scale $\omega_0$to write the energy levels as $E_n = \delta_n\omega_0/2$.
The Hamiltonian may be written as
\begin{equation}
    \mathcal{H}_0^S = \frac{\omega_0}{2} \sum_{n=0}^{N-1} \delta_n\vert e_n \rangle 
    \langle e_n\vert = \frac{\omega_0}{2} \mathcal{H}^{(n)}.
    \label{eq:appfreesystem}
\end{equation}
The diagonal matrix $\mathcal{H}^{(n)}$ represents the spacing of the energy levels, and it 
may describe the spectrum of a $n$-level system, such as qubit $\mathcal{H}^{(2)}=\sigma_3$, 
as well as that of a quantum register of 2 qubits   $\mathcal{H}^{(2,2)} = \left(\sigma_3 \otimes \mathbb{I}_2 + \mathbb{I}_2 \otimes \sigma_3\right)$\cite{reina2002decoherence}. In the second case,  the spectrum might be degenerate. Moreover, where appropriate, we understand the index $n$ as a multiindex $n=(n_1,n_2)$, with each $n_1$,$n_2$ associated respectively with the first qubit and the second qubit.
\par
The reservoir is described by a bath of harmonic oscillator $\mathcal{H}^B_0 = \sum_k \omega_k b^\dag_k b_k$, where $\omega_k$ are the frequencies of the $k$-th bosonic modes. Then, the interaction between the system and the reservoir is given by
\begin{equation}
    \mathcal{H}_I =  \mathcal{H}^{(n)} \otimes \sum_ k(g_k b^\dag_k + g^*_k b_k)
\end{equation}
The quantities $g_k$ are the coupling constants between each energy levels and the $k$-th mode of the bath. We assume they do not depend on the energy level with which they interact. This is justified by the assumption that the system is small compared to the size of the reservoir and a collective interaction is a good approximation. In other words, all the energy levels feel the same local environment. Moreover, we assume that in the case of quantum register, all the qubits interact locally with the same thermal bath \cite{gebbia2020two}.
\par
The model here is exactly solvable.  The evolution of the quantum probe in the interaction 
picture, given the overall system prepared initially in a factorized state $\rho_S \otimes \nu$,
  is given by
\begin{equation}
    \Phi^\beta_t [\rho] = \mathcal{V}^\beta(t ) \circ \mathcal{R}(t) \circ \rho_S\label{redd}
\end{equation}
where the $\circ$ is the Hadamard (entrywise) product and the quantities $\mathcal{V}^\beta$ and 
$\mathcal{R}(t)$ are given by
\begin{equation}
    \mathcal{V}^\beta(t) = \sum_{j,k=0}^{N-1} e^{\frac{(\delta_j-\delta_k)^2}{4}\Gamma(t\vert \beta)} \vert e_j \rangle \langle e_k \vert,
    \label{eq:puredephasing}
\end{equation}
\begin{equation}
    \mathcal{R}(t) = \sum_{j,k=0}^{N-1} e^{i\xi(t)\frac{\delta_j^2-\delta_k^2}{4}} \vert e_j \rangle \langle e_k \vert.
    \label{eq:complexdephasing}
\end{equation}
The evolution for a generic quantum probe, initialized in the state $\rho_S = 
\sum_{jk} \rho_{jk} \vert e_j \rangle \langle e_k \vert $ is given by
\begin{gather}
    \rho_{S\beta}(t) = \sum_{jk} \rho_{jk}\, e^{i\xi(t)\frac{\delta_j^2-\delta_k^2}{4}}\, e^{\frac{(\delta_j-\delta_k)^2}{4} \Gamma(t\vert \beta)}\, \vert e_j \rangle \langle e_k \vert.
    \label{eq:reduceddynamics}
\end{gather}
The functions $\Gamma(t\vert \beta)$ and $\xi(t)$ are defined as follows 
\begin{align}
    \Gamma(t\vert \beta) & = - \sum_k 4 \frac{\vert g_k \vert^2}{\omega_k^2} \left[1-\cos(\omega_k t)\right] \coth\left(\frac{\omega_k\beta}{2}\right) 
    \label{eq:decohfun}
    \\
    \xi(t) & = - 4 \sum_k  \frac{\vert g_k \vert^2}{\omega_k^2} \left[\omega_k t - \sin\left(\omega_k t\right)\right]
    \label{eq:vartildephi}
\end{align}
The first is the decoherence function. It represents the rate of the damping due to the interaction 
and it depends directly on the temperature. The second one is a temperature independent phase function which does not affect the probability of error, since the Hadamard product is distributive.
\par
As final step, we take the continuous limit for the frequency of the bosonic bath, i.e. 
$\sum_k \to \int d\omega f(\omega)$ and $\vert g_k\vert^2 \to \vert g(\omega) \vert^2$, where 
$f(\omega)$ the density of states. Upon definin the spectral density as 
$J(\omega) = 4 f(\omega) \vert g(\omega)\vert^2$, the decoherence function 
\eqref{eq:decohfun} and the temperature-independent phase function \eqref{eq:vartildephi} 
become respectively
\begin{align}
    \Gamma(t\vert \beta) & = - \int^{+\infty}_0 \!\!\! d\omega\, J(\omega) \coth\left(\frac{\omega\beta}{2}\right) \frac{1-\cos(\omega t)}{\omega^2} \\
    \xi(t) & = - \int_0^{+\infty}\!\!\!\! d\omega\,  J(\omega)\, \frac{\omega t - \sin(\omega t)}{\omega^2}
\end{align}
In the following, we consider environments characterized by Ohmic-like spectral densities of the form
\begin{equation}
    J_s(\omega,\omega_c) = \omega_c \left(\frac{\omega}{\omega_c}\right)^s \exp\left(-\frac{\omega}{\omega_c}\right).
\end{equation}
where $\omega_c$ is the cutoff frequency and $s$ is the ohmicity parameter. The cutoff frequency 
is related to the environmental correlation time, and in turn to the decoherence time, whereas 
$s$ sets out the behavior of the spectral density in the low frequency range. Three main 
classes may be identified: the sub Ohmic ($0<s<1$), the Ohmic ($s=1$) and the 
superOhmic ($s>1$)\cite{RevModPhys.59.1,Shnirman_2002}.
\par
Notice that for a dephasing quantum probe the populations of the energy levels are not changed 
by the evolution, which affects only the off diagonal terms of the density matrix of the system.
In other words, there is no exchange of energy between the probe and the system under 
investigation, and the state of the probe is always out-of-equilibrium.
\subsection{\label{subsec:qubit} Out-of-equilibrium qubit probe}
Let us first consider a qubit probe. In this case, $\delta_0 = -1$ and $\delta_1 = +1$ and we make the identification $\vert e_0\rangle \to \vert 0 \rangle$ and $\vert e_1 \rangle \to \vert 1 \rangle$. Thus $\mathcal{R}(t) = \mathbb{I}_2$ and we can write the density matrix \eqref{eq:reduceddynamics} directly in the basis $\vert i\rangle$ at time $t$, obtaining
\begin{equation}
    \rho^\beta_S(t) = \begin{bmatrix}
    1 &  e^{\Gamma(t\vert \beta)} \\
    e^{\Gamma(t\vert \beta)} & 1
    \end{bmatrix} \circ \rho_S
    \label{eq:qubitredstate}
\end{equation}
We assume no a priori information about the two temperatures to be discriminated, 
i.e. $z_1=z_2=\frac12$. The operator $\Lambda$ in Eq. \eqref{eq:Lambda} may be written 
as \begin{align}
    \Lambda =  \begin{pmatrix}
        0 & \rho_{10} \\
        \rho_{01}& 0
    \end{pmatrix}\, \frac{e^{\Gamma (t\vert \beta_1)}-e^{\Gamma (t\vert \beta_2)}}{2}
    \label{eq:lambdanoneq}
\end{align}
where $\rho_{10}$ and $ \rho_{01}$ are the off diagonal elements of the initial state of the probe.
The probability of error is thus given by 
\begin{equation}
    p_{e}^{neq}(\beta_1,\beta_2) =\frac{1}{2}
    \left[1 -  \Big\vert 
    \rho_{10} 
    \left( 
    e^{\Gamma (t\vert \beta_1)} -e^{\Gamma (t\vert \beta_2)} 
    \right)
    \Big\vert \right]
    \label{eq:perrononeq}
\end{equation}
We notice that the probability of error depends only on the off diagonal values of 
the density matrix at time $t=0$ and it does not depend on the value of $\omega_0$. 
Using the Bloch vector formalism, it can be seen that the best preparation is given 
for $\vert \rho_{10} \vert = 1/2$ and $\rho_{00}=\rho_{11}=1/2$. If $\rho_{01}$ is 
real, the optimal probe state is the maximally coherent state $\vert \psi_S (0)
\rangle = 1/\sqrt{2} \left(\vert 0 \rangle +\vert 1 \rangle \right)$. 
\par
Generally speaking, we can exactly find the optimal POVM to be implemented on the probe, which is identified by the projectors of the $\Lambda$ operator \eqref{eq:lambdanoneq}. Indeed, if we write $\rho_{01} = r e^{-i \alpha}$, then we can write the projective measurement in terms of Pauli matrices as
 \begin{align}
     \Pi_1 = \frac{1}{2} \left(\mathbb{I}+ \cos(\alpha) \sigma_x + \sin(\alpha) \sigma_y\right) \\
     \Pi_2 = \frac{1}{2} \left(\mathbb{I}- \cos(\alpha) \sigma_x - \sin(\alpha) \sigma_y\right)
 \end{align}
This is a feasible POVM which depends on the initial preparation but not on the two temperatures
(it does, however, depends on time since the above expression is in the interaction picture).
\par
The above results may be interpreted in terms of coherence of the probe, i.e. the
quantity $\mathcal{C} (\rho) = \sum_{i\neq j} \vert \rho_{ij} \vert$.
For the state in Eq. \eqref{eq:qubitredstate}, the coherence is 
\begin{equation}
     \mathcal{C} (\beta,t) = 2 \vert \rho^0_{01} \vert e^{\Gamma(t\vert \beta)}\,,
 \end{equation} 
and  the probability of error may be written as a function of the coherence only
\begin{equation}
    p_{e}^{neq}(\beta_1,\beta_2) = \frac{1}{2} \left( 1 -  
    \Big \vert \mathcal{C}(\beta_1,t) - \mathcal{C}(\beta_2,t) \Big \vert
    \right)
    \label{eq:perrnoneqcoh}
\end{equation}
Better discrimination is thus obtained for states with 
larger differences of their coherence. In turn, maximally coherent states are optimal 
states, since they are more sensible to decoherence, which is the sole effect of the 
pure dephasing model.
\begin{figure}[h!]
    \centering
    \includegraphics[width=0.99\columnwidth]{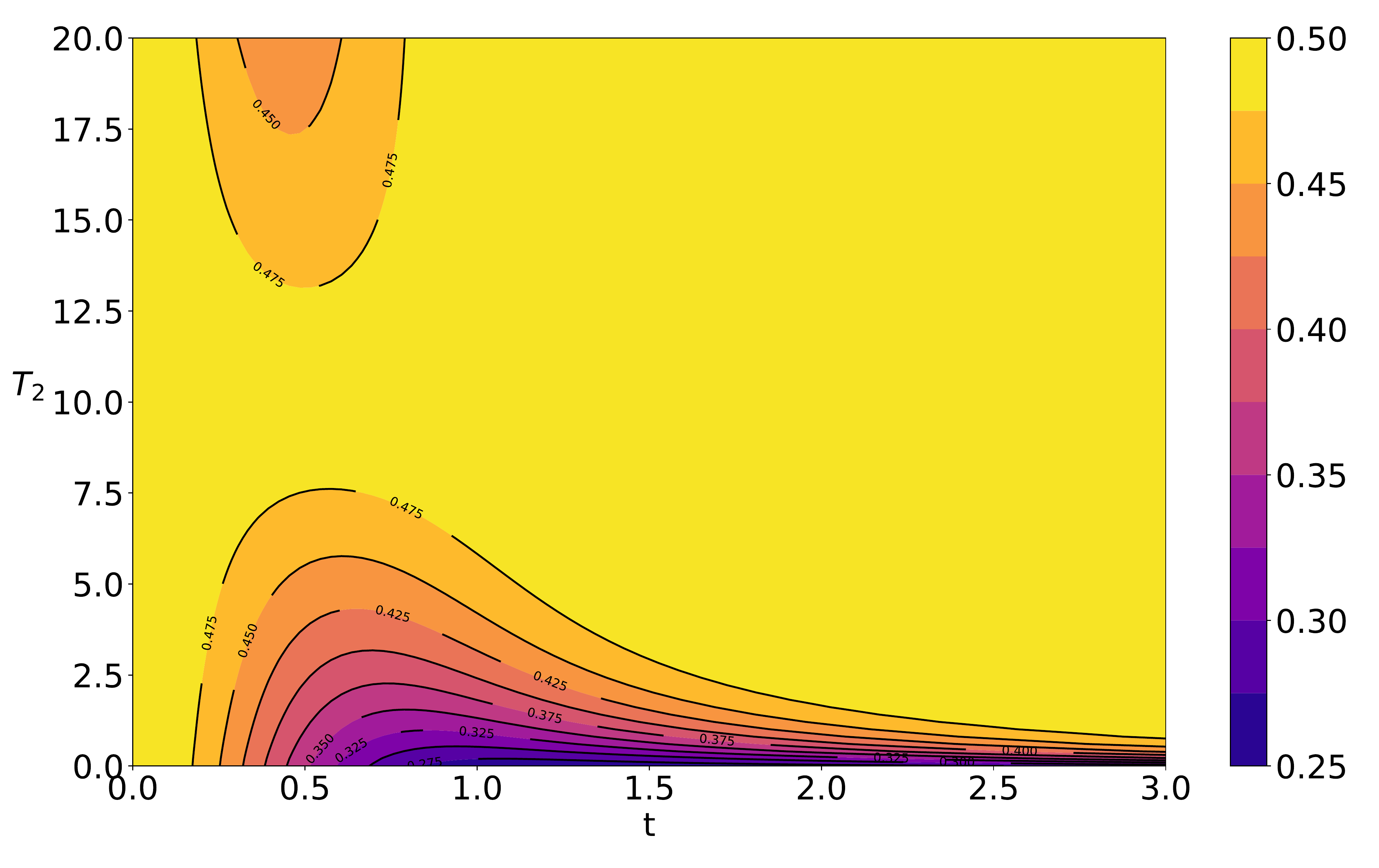}
    \caption{Probability of error $p_{e}^{neq}(T_1,T_2)$ using a dephasing probe, see Eq.  \eqref{eq:perrononeq}, as a function of time $t$ and temperature $T_2$. We set $\vert \rho_{01} \vert = 1/2$, $T_1 = 10$ and we consider an Ohmic environment $s=1$ with a cut-off frequency $\omega_c=1$.}
    \label{plot:noneqhighT}
\end{figure}
\par
In Fig. \ref{plot:noneqhighT} we show the probability of error in Eq. \eqref{eq:perrononeq} 
for an Ohmic environment with fixed  $T_1 = 10$ and we see that for large $T_2$  
it has only small deviations from the maximum error $p_{e}^{neq}=1/2$, meaning that 
in such regime the states are almost indistinguishable. Instead, in the low 
temperature regime and at intermediate time $t$ the $p_{e}^{neq}$ reaches 
smaller values. Similar behaviours may be observed for other values of $s$ 
and $\omega_c$. For this reason, henceforth we will focus only on 
low temperature regimes. 
\par
\begin{figure}[h!]
\centering
\includegraphics[width=0.99\columnwidth]{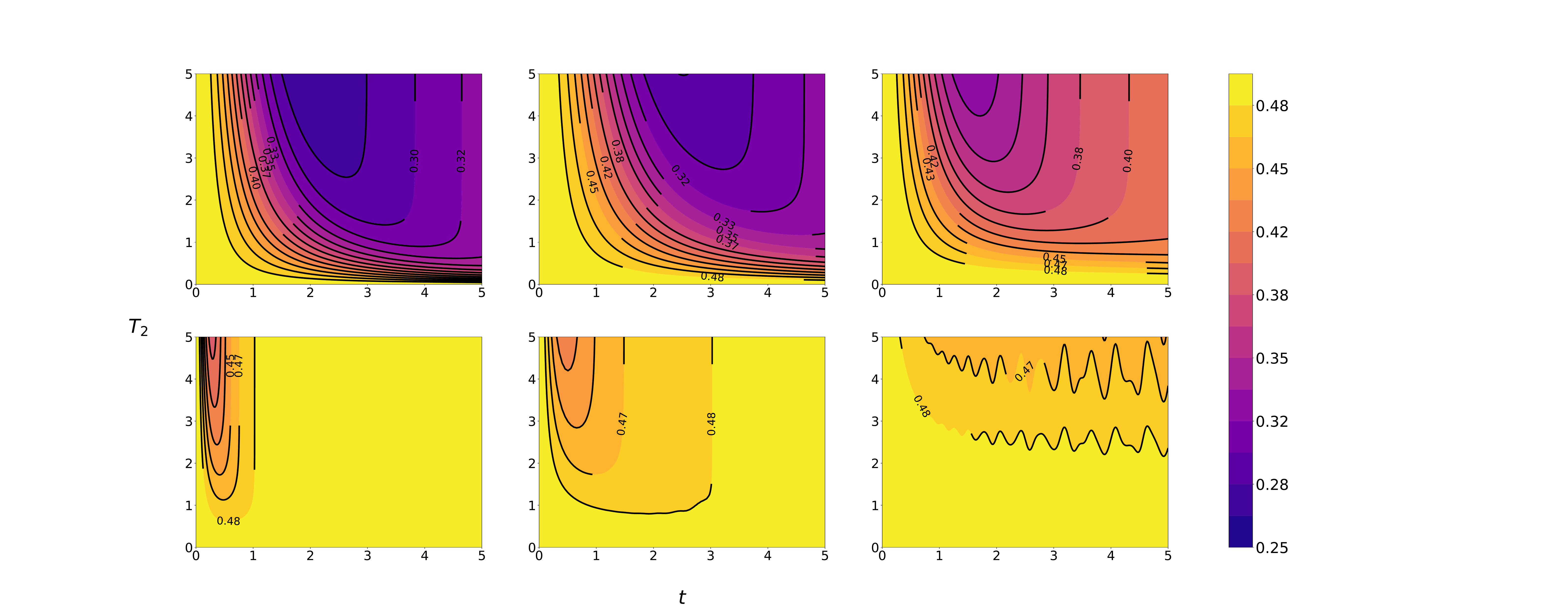}
\caption{Plot of $p_{e}^{neq}(T_1,T_2)$ for a qubit system as a function of time $t$ and of $T_2$ with fixed $T_1 = 0$. We also set $\vert \rho_{01} \vert = 1/2$. First row: $\omega_c = 1/5$; second row: $\omega_c = 5$. First Column $s=0.5$ (subOhmic); second column $s=1$ (Ohmic); third column $s=3$ (superOhmic).}
\label{plot:noneqomegas}
\end{figure}

Next, in Fig. \ref{plot:noneqomegas}, we illustrate how the probability of error depends on 
the ohmicity parameter $s$ and the cut-off frequency $\omega_c$. There, we choose three 
paradigmatic values of $s$ to cover the three Ohmic regimes, that is $s=0.5$, $s=1$, 
and $s=3$. These plots show that a better discrimination is achievable for smaller values 
of the cutoff frequency and in subOhmic environments. This was predictable since in 
superOhmic environments the dephasing effects are smaller and consequently also the 
differences in the coherences are smaller, see \eqref{eq:perrnoneqcoh}. In all the plots, 
we may observe a common behaviour of the the probability of error: at $t=0$ 
$p_{e}^{neq}(T_1,T_2)$ is maximum and then starts to decrease (the exact behaviour 
depending on the nature of the environment) achieving the minimum probability of error. 
After the minimum, there is a slow increase which asymptotically tends the maximum 
error $p_{e}^{neq}$ again. This can be understood by the fact that after a large time, 
the decoherence effects are nearly the same for any values of temperatures $T_1$ 
and $T_2$. 
\par
\begin{figure}[h!]
\centering
\includegraphics[width=0.99\columnwidth]{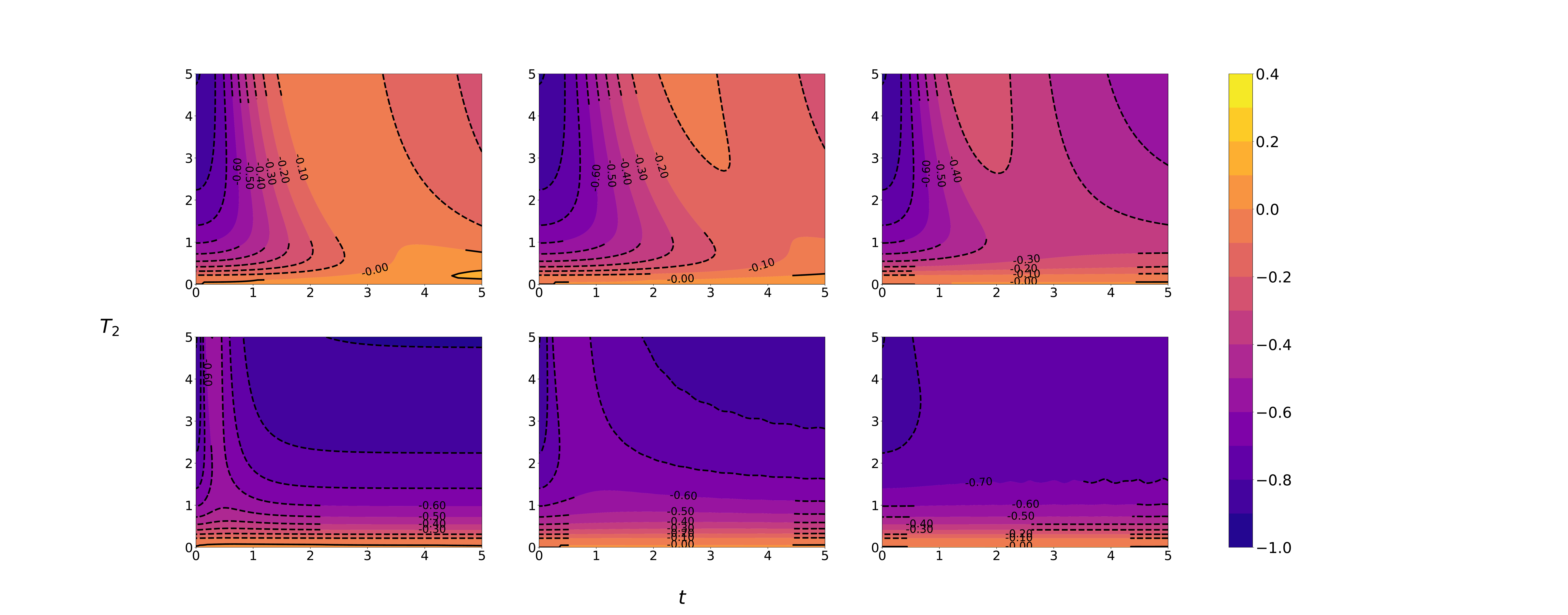}
\includegraphics[width=0.99\columnwidth]{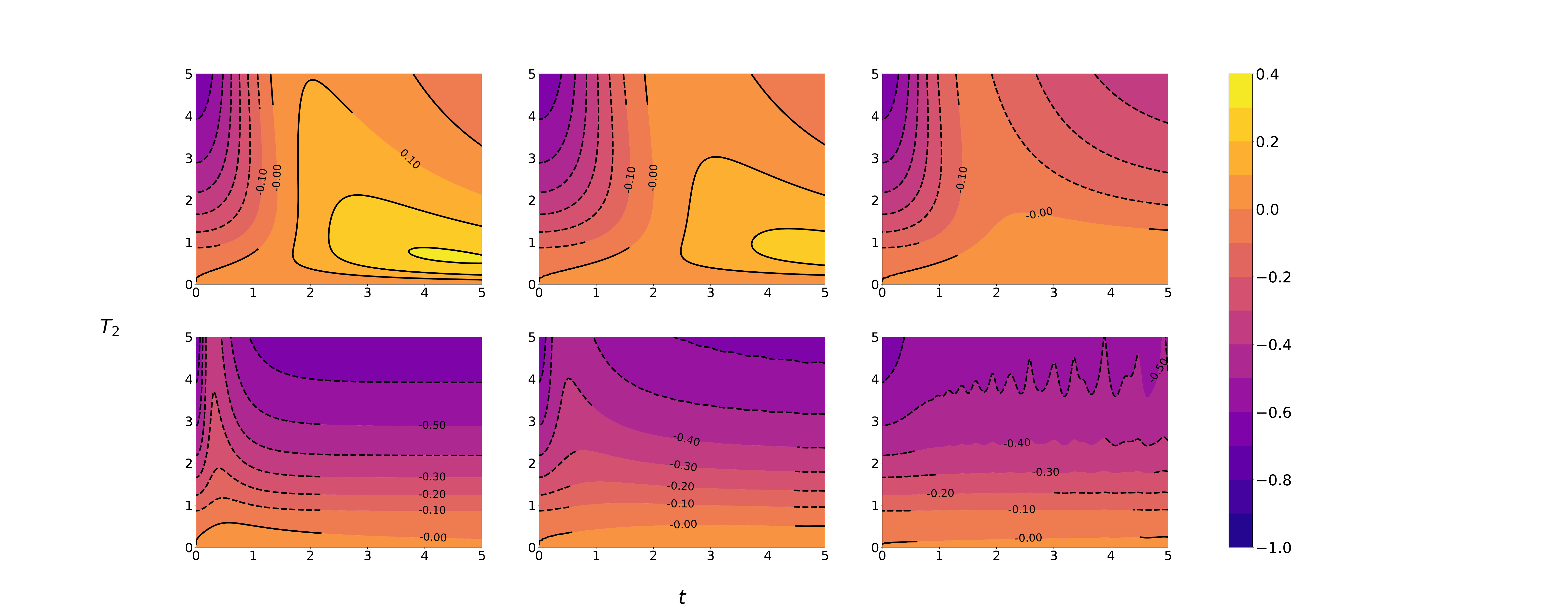}
\caption{Plot of the gain factor $\eta(T_1,T_2)$ \eqref{eq:percentagevarqub} as a function of time $t$ and $T_2$. We set $T_1=0$. The quantum probe is prepared in the maximally coherent state. 
First row:  $\omega_0 = 0.5$ and $\omega_c = 0.2$; second row: $\omega_0 = 0.5$ and $\omega_c = 5.0$; 
third row:  $\omega_0 = 2.0$ and $\omega_c = 0.2$; fourth row: $\omega_0 = 2.0$ and $\omega_c = 5.0$. 
First column $s=0.5$ (subOhmic); second column $s=1$ (Ohmic); third column $s=3$ (superOhmic).}
\label{f:compg}
\end{figure}

Finally, in Fig. \ref{f:compg}, we compare the performance 
of the equilibrium probe studied in Section \ref{sec:thermaleq} 
to those of a dephasing one.
In order to have a faithful comparison, we introduce a gain factor defined as
\begin{equation}
    \eta(T_1,T_2) = 1 - \frac{p_{e}^{neq}(T_1,T_2)}{p_{e}^{eq}(T_1,T_2)}.
    \label{eq:percentagevarqub}
\end{equation}
A positive value of $\eta(T_1,T_2)$ means that the probability of error in the 
non-equilibrium regime is lower than the probability of error in the equilibrium 
regime. Thus $\eta$ quantifies the gain we obtain with a non equilibrium probe. 
On the contrary, negative $\eta$ means that an equilibrium probe provide a lower 
probability of error, and thus the latter is preferable. We show the gain factor 
in Fig. \ref{f:compg} for $T_1=0$ and for different values 
of $s$, $\omega_c$ and $\omega_0$. As it is apparent from the first two rows of 
Fig. \ref{f:compg}, for small values of $\omega_0$ the equilibrium probe is almost everywhere better 
than the dephasing one. This is strongly different from the case of $\omega_0=2$, 
illustrated in the third and fourth rows, where for small values of the cutoff frequency 
(first row) and especially for $s\leq 1$ there is a wide range of interaction times and temperature 
of where the dephasing probe has a lower probability of error. For larger value of the
cutoff frequency (second row), dephasing probes are suitable for very low temperature 
discrimination. 
\subsection{\label{subsec:qutrit} Out-of-equilibrium qutrit probe}
In this section we devote our attention to a $3$-level system with equispaced energy levels. In this system, $\delta_0 = -2$, $\delta_1 = 0$ and $\delta_2 = +2$, and we make the identification $\vert e_0 \rangle \to \vert 0 \rangle$, $\vert e_1 \rangle \to \vert 1 \rangle$ and $\vert e_2 \rangle \to \vert 2 \rangle$. The reduced dynamics is given by Eq. \eqref{redd}, 
with the matrix $\mathcal{V}^\beta(t )$ now given by 
\begin{equation}
    \mathcal{V}^\beta(t ) = \begin{pmatrix}
    1 &  e^{ \Gamma(t\vert \beta)} &  e^{4\Gamma(t \vert \beta)} \\
    e^{\Gamma(t\vert \beta)} & 1 &  e^{\Gamma(t\vert \beta)} \\
    e^{4\Gamma(t\vert \beta)} &  e^{\Gamma(t\vert \beta)} & 1
    \end{pmatrix} 
\end{equation}
In order to compare results with those obtained with a qubit probe, we consider qutrit 
initially prepared in a maximally coherent state $\vert \varphi_3\rangle = (|0\rangle+|1\rangle+|2\rangle)/\sqrt{3}$ and in a qubit-like state $\vert \varphi_3\rangle = (|0\rangle+|2\rangle)/\sqrt{2}$.  
\par
In the first case, the trace  of the operator $\vert \Lambda\vert$ defined 
in \eqref{eq:Lambda} is given by
\begin{align}
    \tr{|\Lambda|} = & \frac{1}{6} \left\vert e^{4\Gamma(t\vert \beta_1)}-e^{4\Gamma(t\vert \beta_2)}\right \vert  \\
    & \times \left[ 1 + \sqrt{1+ 8\left(
    \frac{e^{\Gamma(t\vert \beta_1)}-e^{\Gamma(t\vert \beta_2)}}{e^{4\Gamma(t\vert \beta_1)}-e^{4\Gamma(t\vert \beta_2)}}\right)^2
    }\right]\nonumber
\end{align}
and  the corresponding probability of error $p_{e3}^{neq}(T_1,T_2)$ may be then obtained 
from Eq. \eqref{eq:perrautoval}. For the 
state $|\varphi_q\rangle$,  $p_{e2}^{neq}$ is instead similar to that of the qubit, i.e.
\begin{equation}
  p_{e2}^{neq}(\beta_1,\beta_2) =\frac{1}{2}-\frac{\vert \rho_{10}\vert}{2}\Big\vert e^{4\Gamma (t\vert \beta_1)}-e^{4\Gamma (t\vert \beta_2)} \Big\vert\,,
  \label{eq:perrnoneq4gamma}
\end{equation}
with the difference that the exponentials have a more rapid decrease. The behaviour is thus similar
to that of the qubit, but the minimum is achieved at smaller times.
\par
\begin{figure}[h!]
\centering
   \includegraphics[width=0.9\columnwidth]{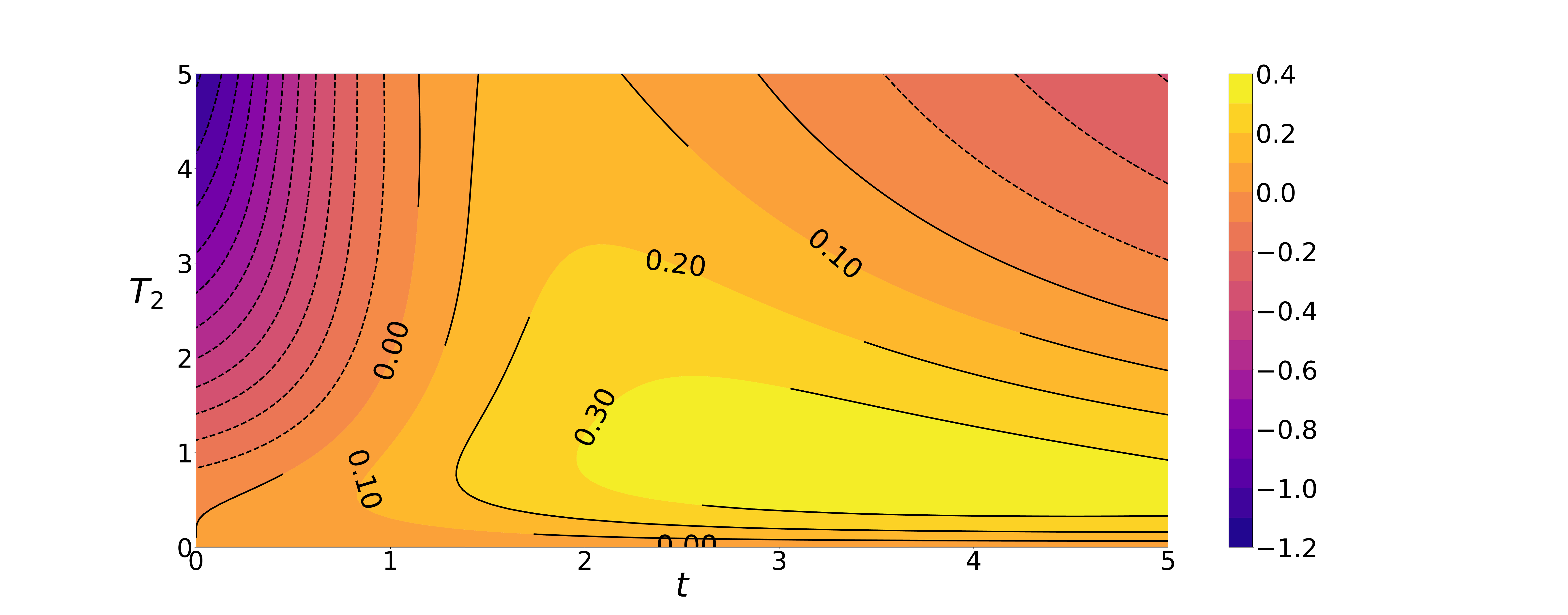}
   \includegraphics[width=0.9\columnwidth]{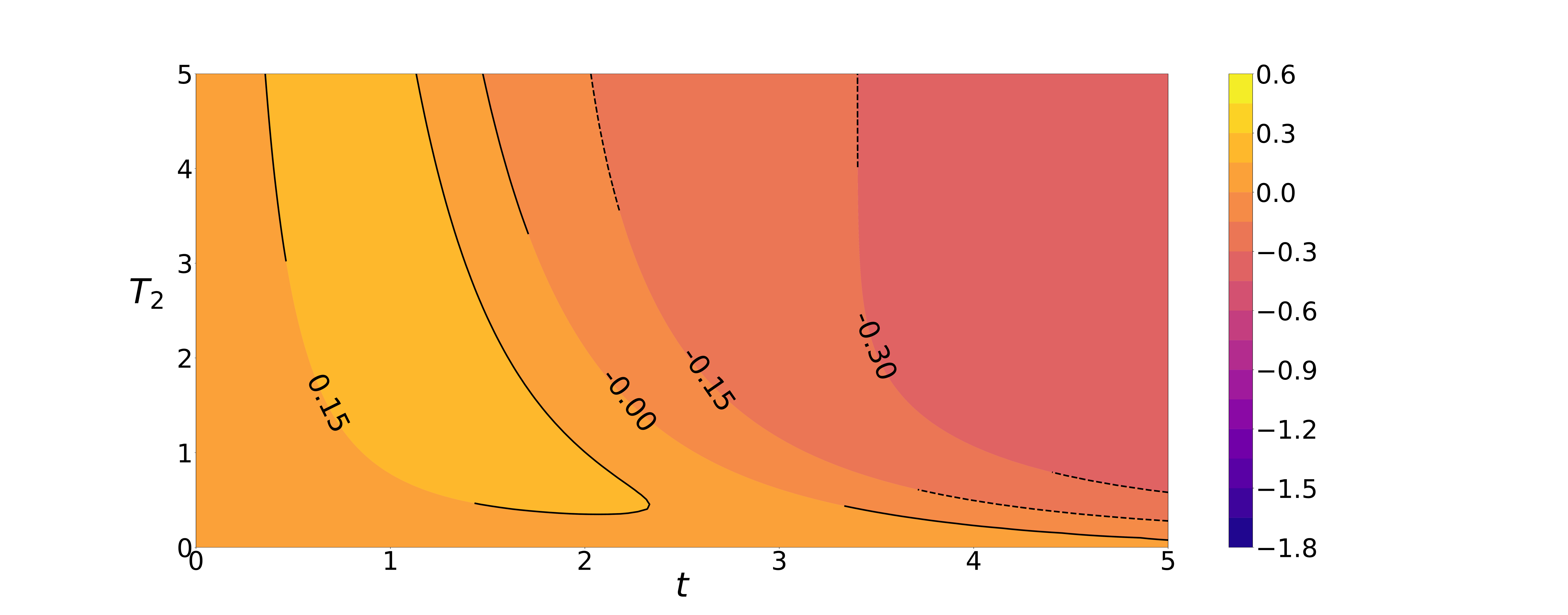}
   \includegraphics[width=0.9\columnwidth]{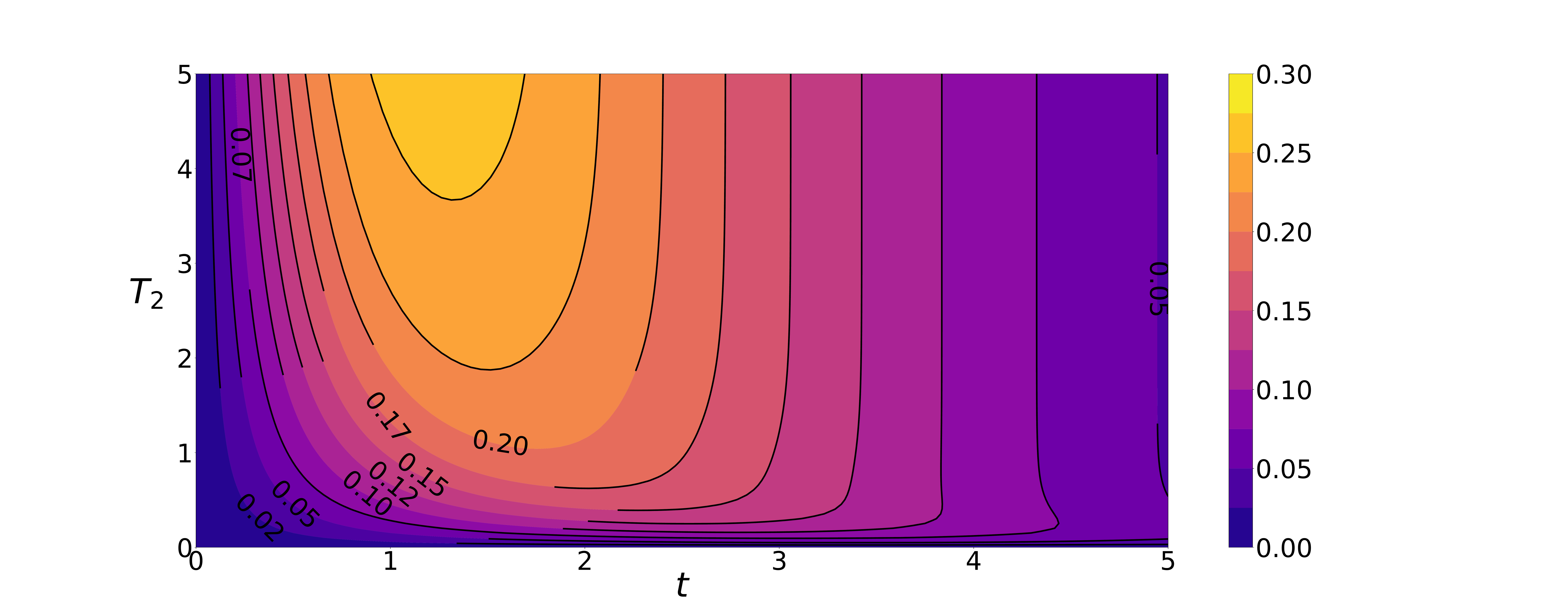}
\caption{Upper panel: the gain factor $\eta_3(0,T_2)$ of Eq. \eqref{eq:comparqutnoneqquteq} as a function of the interaction time anof $T_2$. Middle panel: the gain factor $\eta_2(0,T_2)$ of Eq.\eqref{eq:gainfactqubitvsqutrit4}. Lower panel: the gain factor $\eta_c(0,T_2)$ defined in Eq.\eqref{eq:comparisonqubqut}. We have set $s=0.5$, $\omega_c = 0.2$, $\omega_0 = 2$.}
\label{f:f6}
\end{figure}

We compare now the different probes. At first, we compare the probability of error 
obtained with an equilibrium probe to that obtained for a qutrit initially 
prepared in a maximally coherent state. The gain factor is defined as 
\begin{equation}
    \eta_3(T_1,T_2) = 1- \frac{p_{e3}^{neq}(T_1,T_2)}{p_{e3}^{eq}(T_1,T_2)}\,,
    \label{eq:comparqutnoneqquteq}
\end{equation}
and the value for $T_1=0$ in shown in the upper panel of Fig. \ref{f:f6}. We see a positive 
gain factor for intermediate values of the interaction time low values of $T_2$. 
\par
We then compare the performance of a qutrit prepared in the state $|\varphi_2\rangle$ with 
that of a maximally coherent qubit by means the gain factor 
\begin{equation}
    \eta_2(T_1,T_2) = 1- \frac{\tilde{p}_{e2}^{neq}(T_1,T_2)}{p_{e}^{neq}(T_1,T_2)}\,,
    \label{eq:gainfactqubitvsqutrit4}
\end{equation}
where $p_{e}^{neq}(T_1,T_2)$ is given in Eq. (\ref{eq:perrononeq}).
We show $\eta_2(0,T_2)$ in the middle panel of \ref{f:f6}. We see that for short interaction 
times the qutrit state has always a lower probability of error for all temperatures $T_2$. 
Moreover, the lower is $T_2$ the larger is the time for which $\eta > 0 $. 
\par
Finally, we compare the probabilities of error obtained with a maximally coherent qubit 
and a maximally coherent qutrit. The corresponding gain factor is defined as
\begin{equation}
    \eta_c(T_1,T_2) = 1- \frac{p_{e3}^{neq}(T_1,T_2)}{p_{e}^{neq}(T_1,T_2)}
    \label{eq:comparisonqubqut}
\end{equation}
and it is shown in the lower panel of Fig. \ref{f:f6}. We see that the gain factor 
is always positive. However, the largest improvement is achievable in the early phase
 of the interaction and for larger values of $T_2$. We thus conclude that a qutrit 
 probe allows one to achiee lower error probability using a shorter interaction time.
Upon comparing the different panels of Fig. \ref{f:f6} we see that preparing the qutrit 
in the initial state $|\varphi_2\rangle$ is suitable to discriminate low temperatures, whereas 
the maximally coherent preparation may be better employed when the two temperatures 
are more different.
\par
\subsection{Out-of-equilibrium quantum register made of two qubits}
Finally, we investigate the performance of a quantum register of two qubits interacting 
locally with the thermal bath \cite{gebbia2020two,reina2002decoherence}. In this case 
the matrix of the levels spacing is given by
\begin{equation}
    \mathcal{H}^{(2,2)} = \left(\sigma_3 \otimes \mathbb{I}_2 + \mathbb{I}_2 \otimes \sigma_3\right)
\end{equation}
We thus obtain that $\delta_{00} = -2$, $\delta_{01} = 0 = \delta_{10}$ and $\delta_{11} = +2$ 
and make the identifications $\vert e_0 \rangle \to \vert 00\rangle$, $\vert e_1 \rangle \to \vert 01\rangle$, $\vert e_2 \rangle \to \vert 10\rangle$ and $\vert e_3 \rangle \to \vert 11 \rangle$. 
The reduced dynamics is given by Eq. \eqref{redd}, 
with the matrix $\mathcal{V}^\beta(t )$ now given by 
\begin{equation}
    \mathcal{V}^\beta(t ) =
 \begin{pmatrix}
    1   &   e^{\Gamma(t\vert \beta)}   &   e^{\Gamma(t\vert \beta)}    &   e^{4\Gamma(t\vert \beta)}  \\
    e^{\Gamma(t\vert \beta)}    &   1   &   1   &   e^{\Gamma(t\vert \beta)}    \\  
    e^{\Gamma(t\vert \beta)}    &   1   &   1   &   e^{\Gamma(t\vert \beta)}    \\
    e^{4\Gamma(t\vert \beta)}   &   e^{\Gamma(t\vert \beta)}    &   e^{\Gamma(t\vert \beta)}    &   1
    \end{pmatrix}
    \label{eq:dynamicsQR}
\end{equation}
For the register initially prepared in a maximally coherent state $|\varphi_4\rangle = \frac14 
\sum_k |e_k\rangle$ we have 
\begin{align}
    \tr{|\Lambda|} = & \frac{1}{8} \left\vert e^{4\Gamma(t\vert \beta_1)}-e^{4\Gamma(t\vert \beta_2)}\right \vert  \\
    & \times \left[ 1 + \sqrt{1+ 16\left(
    \frac{e^{\Gamma(t\vert \beta_1)}-e^{\Gamma(t\vert \beta_2)}}{e^{4\Gamma(t\vert \beta_1)}-e^{4\Gamma(t\vert \beta_2)}}\right)^2
    }\right]\nonumber
\end{align}
and  the corresponding probability of error $p_{e4}^{neq}(T_1,T_2)$ may be then obtained 
from Eq. \eqref{eq:perrautoval}. Having at disposal two qubits, one may wonder whether 
entanglement may play a role in the discrimination task. We thus consider the four Bell states as possibile initial preparation of the probe register. As it may be easily seen, the states 
$\vert \Psi^\pm \rangle = 1/\sqrt{2} (\vert 01 \rangle \pm \vert 10 \rangle)$ are useless since they are invariant under dynamics \ref{eq:dynamicsQR}. Concerning the states $\vert \Phi^\pm \rangle = 1/\sqrt{2}(\vert 00 \rangle \pm \vert 11 \rangle)$ the  probability of error
is equal to that of the qutrit \eqref{eq:perrnoneq4gamma}.
\par
\begin{figure}[h!]
\centering
  \includegraphics[width=0.95\columnwidth]{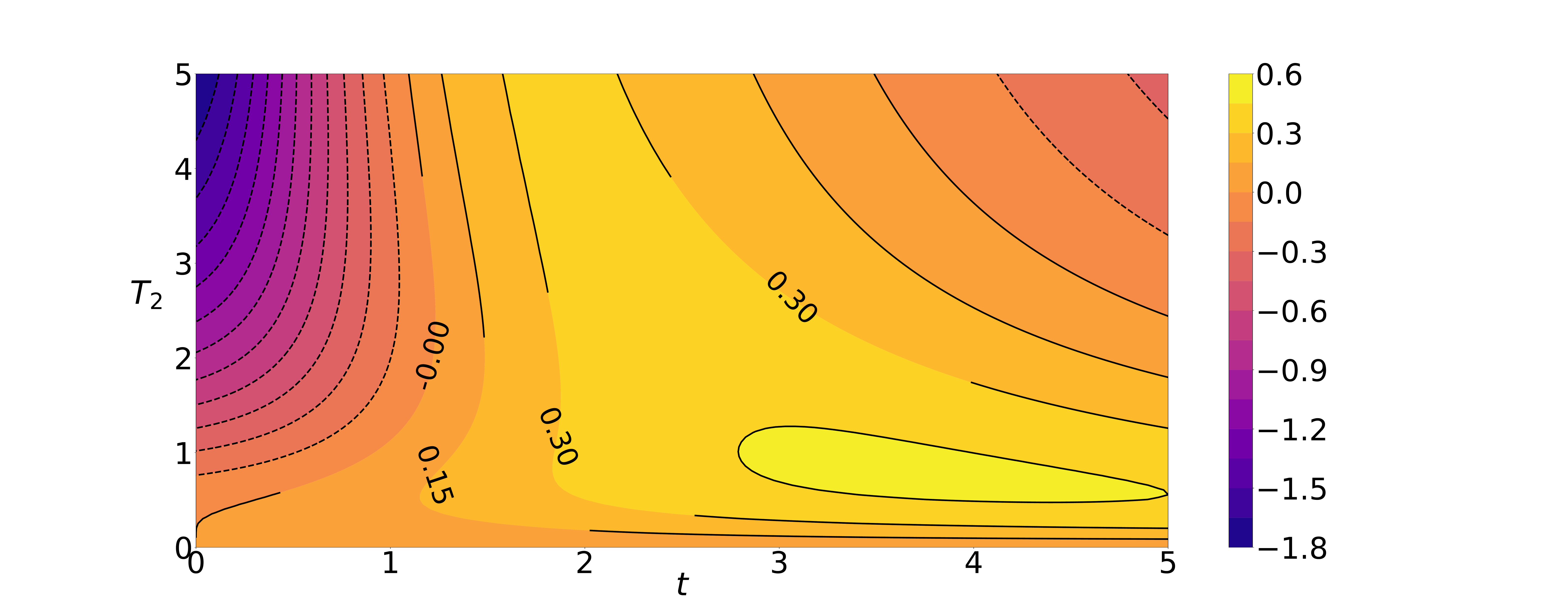}
  \includegraphics[width=0.95\columnwidth]{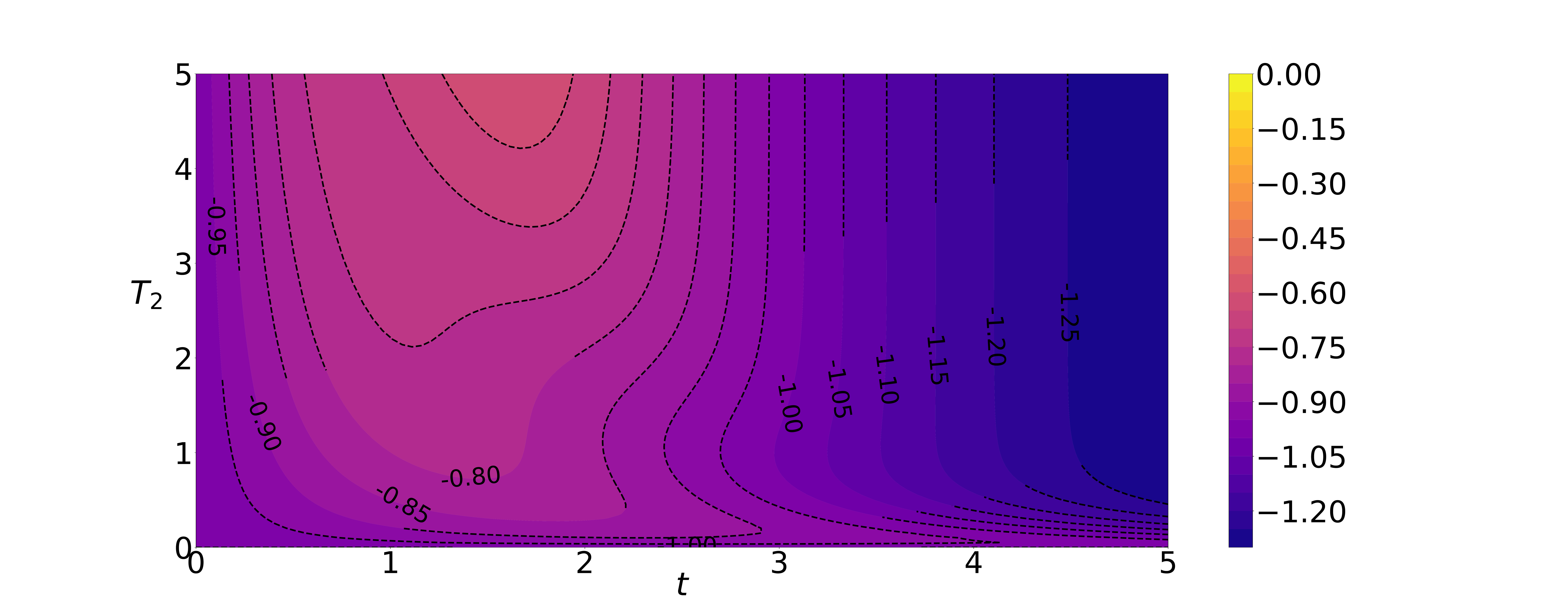}
\caption{Upper panel: the gain factor $\eta_4(0,T_2)$ defined in Eq. \eqref{eq:eta1QR2} as
a function of the interaction time and $T_2$. Lower panel: the gain factor $\eta_{42}(0,T_2)$
defined in Eq. \eqref{eq:eta2QR2}. We set $s=0.5$, $\omega_c = 0.2$, $\omega_0 = 2$. }
\label{f:f7}
\end{figure}

In order to asses the quantum register as quantum probe we first define the gain factor 
\begin{equation}
    \eta_4(T_1,T_2) = 1-\frac{p^{neq}_{e4}(T_1,T_2)}{p^{eq}_{e4}(T_1,T_2)}
    \label{eq:eta1QR2}
\end{equation}
to compare the performance of the two-qubit dephasing probe with the corresponding equlibrium one.
We show $\eta_4(0,T_2)$ in the upper panel of \ref{f:f7}, where we observe a behaviou similar to 
the previous dephasing probes. Here, the gain factor is larger for lower values of $T_2$ and for intermediate times. 
\par
It is also of interest to compare the performance of a (maximally coherent) 
two-qubit probe with that of a scheme using two consecutive and independent (maximally coherent) 
single-qubit probes. The corresponding figure of merit is given by the gain factor
\begin{equation}
    \eta_{42}(T_1,T_2) = 1 - \frac{p^{neq}_{e4}(T_1,T_2)}{\left[p^{neq}_{e}(T_1,T_2)\right]^2}\,,
    \label{eq:eta2QR2}
\end{equation}
where $p_{e}^{neq}(T_1,T_2)$ is given in Eq. (\ref{eq:perrononeq}).
As it is apparent from the lower panel of Fig. \ref{f:f7}, there is no advantage in 
using two qubits simultaneously since the gain factor is always negative in the considered 
region. 
\section{\label{subsec:comparison} Conclusion}
In this paper, we have analyzed in details the use of quantum probes to discriminate 
two structured baths at different temperatures. In particular, we have addressed 
quantum probes interacting with the thermal bath by a dephasing Hamiltonian and 
compared the discrimination performance with those of equilibrium probes.
\par
At first, we have addressed the discrimination problem for an equilibrium probe and
evaluated the probability of error in the discrimination problem, showing 
that energy measurement is optimal in this regime . We have then moved to 
out-of-equilibrium dephasing probes and derived the exact reduced dynamics for a finite 
quantum system locally interacting with a Ohmic-like thermal bath. Upon exploiting this result,  
we have studied the behaviour of the probability of error as a function of 
the interaction time and found that in the low-temperature regime 
out-of-equilibrium probes outperform  equilibrium ones. For all the environments here considered
there is a finite value of the interaction time minimizing the probability of error. In particular,  for sub-Ohmic environments decoherence effects are stronger, and this leads to lower error 
probability for shorter interaction times.  In turn, it results that for qubits systems, 
maximally coherent states represent the best preparation of the probe for the discrimination 
task.
\par
We have also compared qubit probes with qutrit ones, and have shown numerically that 
qutrits allows one to achieve lower error probability. Finally, we have also compared
schemes based on simultaneous probes (i.e. quantum registers made of two qubits) to those 
of based on two single-qubit probes, showing that the latter has always the lowest 
probability of error. Overall, our results indicate 
that dephasing quantum probes are useful for the taks of discriminating temperatures, 
and that out-of-equilibrium coherent quantum probes represent a resource not only for 
quantum estimation but also for quantum discrimination.
\acknowledgments
The authors thank C. Benedetti, M. Bina and S. Razavian for useful discussions. 
MGAP is member of INdAM.

\appendix

\bibliography{biblio.bib}
\end{document}